\documentclass{llncs}

\usepackage{amsmath}
\usepackage{amsfonts}
\usepackage{amssymb}
\usepackage{verbatim}
\usepackage{thanks}
\usepackage{latexsym}
\usepackage{times}

\def \integer   {{\mathbb Z}}

\newcommand{\IF}{\mbox{{\bf if}\ }}
\newcommand{\FI}{\mbox{{\bf fi}}}
\newcommand{\DO}{\mbox{{\bf do}\ }}
\newcommand{\OD}{\mbox{{\bf od}}}
\newcommand{\WHILE}{\mbox{{\bf while}\ }}
\newcommand{\END}{\mbox{{\bf end}}}
\newcommand{\THEN}{\mbox{\ {\bf then}\ }}
\newcommand{\ELSE}{\mbox{\ {\bf else}\ }}
\newcommand{\T}{\mbox{{\bf true}}}
\newcommand{\F}{\mbox{{\bf false}}}

\newcommand{\ES}{\mbox{$\emptyset$}}

\newcommand{\BEGIN}{\mbox{{\bf begin}}}
\newcommand{\block}[1]{\mbox{$\BEGIN \ {#1}\ \END$}}
\newcommand{\local}{\mbox{{\bf local}\ }}

\newcommand{\ra}{\mbox{$\:\rightarrow\:$}}
\newcommand{\tra}{\mbox{$\:\rightarrow^*\:$}}

\newcommand{\A}{\mbox{$\ \wedge\ $}}

\newcommand{\Or}{\mbox{$\ \vee\ $}}

\newcommand{\U}{\mbox{$\:\cup\:$}}
\newcommand{\I}{\mbox{$\:\cap\:$}}
\newcommand{\sse}{\mbox{$\:\subseteq\:$}}

\newcommand{\Mo}{\mbox{$\:\models\ $}}

\newcommand{\Mt}{\mbox{$\:\models_{\it tot}\ $}}

\newcommand{\fa}{\mbox{$\forall$}}
\newcommand{\te}{\mbox{$\exists$}}

\newcommand{\LL}{\mbox{$\ldots$}}

\newcommand{\MS}[1]{\mbox{${\cal M}[\![{#1}]\!]$}}

\newcommand{\MT}[1]{\mbox{${\cal M}_{{\it tot}}[\![{#1}]\!]$}}

\newcommand{\ITE}[3]{\mbox{$\IF {#1} \THEN {#2} \ELSE {#3}\ \FI$}}
\newcommand{\IT}[2]{\mbox{$\IF {#1} \THEN {#2}\ \FI$}}

\newcommand{\WDD}[2]{\mbox{$\WHILE {#1}\ \DO {#2}\ \OD$}}

\newcommand{\HT}[3]{\mbox{$\{{#1}\}\ {#2}\ \{{#3}\}$}}

\newcommand{\sg}{{\mbox{$\sigma$}}}

\newcommand{\B}[1]{\mbox{$[\![{#1}]\!]$}}       
\newcommand{\C}[1]{\mbox{$\{{#1}\}$}}           

\newcommand{\NI}{\noindent}
\newcommand{\HB}{\hfill{$\Box$}}
\newcommand{\VV}{\vspace{5 mm}}
\newcommand{\III}{\vspace{3 mm}}

\newenvironment{mytabbing}{\begin{tabbing}\hspace*{\leftskip}\=\+\kill}{\end{tabbing}}

\newcommand{\szkew}[1]{\relax \setbox0=\hbox{\kern -24pt $\displaystyle#1$\kern 0pt }%
\box0}
{\catcode`\@=11 \global\let\ifjusthvtest@=\iffalse}

\newcounter{oldmycaption}

\def\mylabel{\label}
\def\myref{\ref}

\newcommand{\m}{m}

\newcommand{\n}{n}
\newcommand{\p}{pi}
\newcommand{\lf}{le}
\newcommand{\ri}{ri}

\def\nlni{\par\ifvmode\removelastskip\fi\vskip\baselineskip\noindent}

\newcommand{\mynewrule}[4]{
   \newenvironment{#1}{\nlni\begingroup
                       \refstepcounter{#3}
                       {\bf #2 \csname the#3\endcsname#4}
                      }
                      {\endgroup\vskip\baselineskip}
   }
\newcounter{RuleCnt}[chapter]
\mynewrule{Rule}{RULE}{RuleCnt}{:}
\mynewrule{Axiom}{AXIOM}{RuleCnt}{:}

\newcounter{AuxRuleCnt}
\mynewrule{AuxRule}{RULE}{AuxRuleCnt}{:}
\mynewrule{AuxAxiom}{AXIOM}{AuxRuleCnt}{:}

\newenvironment{Proof}
      {\medskip\noindent{\bf Proof.}}
      {\hfill$\Box$\medskip}

\def\smallromani{\renewcommand{\theenumi}{\roman{enumi}}
\renewcommand{\labelenumi}{(\theenumi)}}

\begin{document}

\author{Krzysztof R. Apt\inst{1,2} \and Frank S. de Boer\inst{1,3} \and Ernst-R\"{u}diger Olderog\inst{4}}

\institute{Centre for Mathematics and Computer Science (CWI), Amsterdam, The Netherlands 
\and University of Amsterdam, Institute of Language, Logic and Computation, Amsterdam
\and Leiden Institute of Advanced Computer Science, University of Leiden, The Netherlands
\and Department of Computing Science, University of Oldenburg, Germany
}

\title{Modular Verification of Recursive Programs}

\maketitle
\begin{abstract}
We argue that verification of recursive programs by means of the assertional method of C.A.R. Hoare
can be conceptually simplified using a modular reasoning. In this approach some properties of the program are established
first and subsequently used to establish other program properties.
We illustrate this approach by providing a modular correctness proof of the $Quicksort$ program.
\end{abstract}

\section{Introduction}
\label{sec:intro}

Program verification by means of the assertional method of Hoare
(so-called Hoare's logic) is by now well-understood.  One of its
drawbacks is that it calls for a
tedious manipulation of assertions, which is error prone. The support
offered by the available by interactive proof checkers, such as PVS
(Prototype Verification System), see \cite{OS03}, is very limited.

One way to reduce the complexity of an assertional correctness proof is by
organizing it into a series of simpler proofs.  For example, to prove
$\HT{p}{S}{q_1 \A q_2}$ we could establish $\HT{p}{S}{q_1}$ and
$\HT{p}{S}{q_2}$ separately (and \emph{independently}). Such an
obvious approach is clearly of very limited use.

In this paper we propose a different approach that is appropriate for
recursive programs. In this approach a simpler property, say
$\HT{p_1}{S}{q_1}$, is established first and then \emph{used in the
  proof} of another property, say $\HT{p_2}{S}{q_2}$. This allows us
to establish $\HT{p_1 \A p_2}{S}{q_1 \A q_2}$ in a modular way.  It is
obvious how to generalize this approach to an arbitrary sequence of
program properties for which the earlier properties are used in the
proofs of the latter ones. So, in contrast to the simplistic approach
mentioned above, the proofs of the program properties are \emph{not}
independent but are arranged instead into an acyclic directed
graph.

We illustrate this approach by providing a modular correctness proof
of the $Quick$-$sort$ program due to \cite{Hoa62}.  This yields a
correctness proof that is better structured and conceptually easier to
understand than the original one, given in \cite{FH71}.  A minor point
is that we use different proof rules concerning procedure calls and
also provide an assertional proof of termination of the program, a
property not considered in \cite{FH71}.  (It should be noted that
termination of recursive procedures with parameters within the
framework of the assertional method was considered only in the
eighties, see, e.g., \cite{Apt81b}. In these proofs some subtleties
arise that necessitate a careful exposition, see \cite{AB90}.)

We should mention here two other references concerning formal
verification of the $Quicksort$ program.  In \cite{FM99} the proof of
$Quicksort$ is certified using the interactive theorem prover Coq,
while in \cite{Kal90} a correctness proof of a non-recursive version
of $Quicksort$ is given.

The paper is organized as follows.  In the next section we introduce a
small programming language that involves recursive procedures with
parameters called by value and discuss its operational semantics.
Then, in Section \ref{sec:ver} we introduce a proof system for proving
partial and total correctness of these programs.  The presentation in
these two sections is pretty standard except for the treatment of the
call-by-value parameter mechanism that avoids the use of substitution.

Next, in Section \ref{sec:mod} we discuss how the correctness proofs,
both of partial and of total correctness, can be structured in a
modular way.  In Section \ref{sec:quick} we illustrate this approach
by proving correctness of the $Quicksort$ program while in Section \ref{sec:conc}
we discuss related work and draw some conclusions. Finally, in the
appendix we list the used axioms and proof rules concerned with
non-recursive programs.  The soundness of the considered proof systems
is rigorously established in \cite{ABO09} using the operational
semantics of \cite{Plo81,Plo04}.

\section{A small programming language}
\label{sec:pro}

\subsection*{Syntax} 

We use \emph{simple} variables and \emph{array} variables. Simple
variables are of a basic type (for example \textbf{integer} or
\textbf{Boolean}), while array variables are of a higher type (for
example $\textbf{integer} \times \textbf{Boolean} \ra \textbf{integer}$). A
\emph{subscripted variable} derived from an array variable $a$ of type $T_1
\times \LL \times T_n \ra T$ is an expression of the form $a[t_1, \LL,
t_n]$, where each expression $t_i$ is of type $T_i$.

In this section we introduce a class of recursive programs
as an extension of the class of {\bf while} programs which are generated by the 
following grammar:
\[
 S::=skip \mid u:=t \mid \bar{x}:=\bar{t} \mid S_1;\ S_2 \mid 
     \ITE{B}{S_1}{S_2} \mid \WDD{B}{S_1},
\]
where $S$ stands for a typical statement or program, $u$ for a simple
or subscripted variable, $t$ for an expression (of the same type as
$u$), and $B$ for a Boolean expression.  Further, $\bar{x}:=\bar{t}$
is a parallel assignment, with $\bar{x} = x_1,\dots,x_n$ a non-empty
list of distinct simple variables and $\bar{t}=t_1,\dots,t_n$ a list
of expressions of the corresponding types.  The parallel assignment
plays a crucial role in our modelling of the parameter passing.  We do not
discuss the types and only assume that the set of basic types includes
at least the types \textbf{integer} and \textbf{Boolean}.
As an abbreviation we introduce
$\IT{B}{S} \equiv \ITE{B}{S}{skip}$.

Given an expression $t$, we denote by $var(t)$ the set of all simple
and array variables that appear in $t$.  Analogously, given a program
$S$, we denote by $var(S)$ the set of all simple and array variables
that appear in $S$, and by $change(S)$ the set of all simple and array
variables that can be modified by $S$, i.e., the set of variables that
appear on the left-hand side of an assignment in $S$.

We arrive at recursive programs by adding recursive procedures with
call-by-value parameters. To distinguish between local and global
variables, we first introduce a \emph{block statement} by the grammar
rule
\[
  S::= \block{\local \bar{x}:=\bar{t}; S_1}.
\]
A block statement introduces a non-empty sequence $\bar{x}$ of simple local
variables, all of which are explicitly initialized by means of a
parallel assignment $\bar{x}:=\bar{t}$, 
and provides an explicit scope for these simple local variables.  
The precise explanation of a scope is more complicated because the
block statements can be nested.  

Assuming $\bar{x} = x_1, \LL, x_k$ and $\bar{t} = t_1, \LL, t_k$, each
occurrence of a local variable $x_i$ within the statement $S_1$
\emph{and not} within another block statement that is a subprogram of
$S_1$ refers to the same variable.  Each such variable $x_i$ is
initialized to the expression $t_i$ by means of the parallel
assignment $\bar{x} := \bar{t}$.  Further, given a statement $S'$ such
that $\block{\local \bar{x}:=\bar{t}; S_1}$ is a subprogram of $S'$,
all occurrences of $x_i$ in $S'$ outside this block statement refer to
some other variable(s).

Procedure calls with parameters are introduced by the grammar rule
\[
  S::= P(t_1,\ldots,t_n),
\]
where $P$ is a procedure identifier and $t_1,\ldots,t_n$, 
with $n \geq 0$, are expressions called \emph{actual parameters}. 
The statement $P(t_1,\ldots,t_n)$ is called a \emph{procedure call}.
The resulting class of programs is then called \emph{recursive programs}.

Procedures are defined by \emph{declarations} of the form
\[
 P(u_1,\ldots,u_n)::S,
\]
where $u_1,\ldots,u_n$ are distinct simple variables, called 
\emph{formal parameters} of the procedure $P$ and $S$ is the \emph{body} of the procedure $P$.

We assume a given set of procedure declarations $D$ such that each procedure that appears in $D$ has a unique declaration in $D$. 
When considering recursive programs
we assume that all procedures whose calls
appear in the considered recursive programs are declared in $D$.
Additionally, we assume that the procedure calls are
\emph{well-typed}, which means that the numbers of formal and actual
parameters agree and that for each parameter position the types of the
corresponding actual and formal parameters coincide.

Given a recursive program $S$, we call a variable $x_i$ \emph{local}
if it appears within a subprogram of $D$ or $S$ of the form
$\block{\local \bar{x}:=\bar{t}; S_1}$ with $\bar{x} = x_1, \LL, x_k$,
and \emph{global} otherwise.

To avoid possible name clashes between local and global variables
we assume that given a set of procedure declarations $D$ and a recursive
program $S$, no local variable of $S$ occurs in $D$. So
given the procedure declaration
\[
P::\ITE{x=1}{b:=\T}{b:=\F}
\]
the program 
\[
S \equiv \block{\local  x:=1; P}
\]
is not allowed. If it were, the semantics we are about to introduce
would allow us to conclude that $\HT{x=0}{S}{b}$ holds.  However,
the customary semantics of the programs in the presence of procedures
prescribes that in this case $\HT{x=0}{S}{\neg b}$ should hold, as the
meaning of a program should not depend on the choice of the names of
its local variables. (This is a consequence of the so-called \emph{static
scope} of the variables that we assume here.)

This problem is trivially solved by just renaming the `offensive'
local variables to avoid name clashes, so by considering here the
program $\block{\local y:=1; P}$ instead of $S$.  Once we limit ourselves to
recursive programs no local variable of which occurs in the considered
set of procedure declarations, the semantics we introduce ensures that
the names of local variables indeed do not matter. More precisely, the
programs that only differ in the choice of the names of local
variables and obey the above syntactic restriction have then identical
meaning. In what follows, when considering a recursive program $S$ in
the context of a set of procedure declarations $D$ we always
implicitly assume that the above syntactic restriction is satisfied.

The local and global variables play an analogous role to the bound and
free variables in first-order formulas or in $\lambda$-terms. In fact,
the above syntactic restriction corresponds to the `Variable
Convention' of \cite[page 26]{Bar84} according to which ``all bound
variables are chosen to be different from the free variables.''

Note that the above definition of programs puts no restrictions 
on the actual parameters in procedure calls; in particular they
can be formal parameters or global variables.

\subsection*{Semantics}

For recursive programs we use 
a structural operational semantics in the sense of Plotkin \cite{Plo04}.
As usual, it is defined in terms of transitions between configurations.
A \emph{configuration} $C$ is a pair $< S,\ \sg >$ consisting a statement $S$ 
that is to be executed and a state $\sg$ that assigns a value 
to each variable (including local variables).
A \emph{transition} is written as a step $C \ra C'$ between configurations.
To express termination we use the empty statement $E$; a configuration
$< E,\ \sg >$ denotes termination in the state $\sg$.

Transitions are specified by the transition axioms and rules
which are defined in the context of a set $D$ of procedure declarations.
The only transition axioms that are somewhat non-standard are the ones that
deal with the block statement and the procedure calls, 
in that they avoid the use of substitution thanks to the use of parallel assignment:

\begin{mytabbing}
\quad \= $<\block{\local \bar{x}:=\bar{t}; S},\sigma>\ra <\bar{x}:=\bar{t}; S;\bar{x}:=\sigma(\bar{x}),\sigma>$,
\end{mytabbing}

\begin{mytabbing}
\quad \= $<P(\bar{t});R,\sigma> \ra 
          <\block{\local \bar{u}:=\bar{t};S};R,\sigma>$, \\[2mm]
      \> where $P(\bar{u}) ::S\in D$.
\end{mytabbing}

The first axiom ensures that the local variables are initialized as
prescribed by the parallel assignment and that upon termination the
global variables whose names coincide with the local variables are
restored to their initial values, held at the beginning of the block
statement. This is a way of implicitly modeling a \emph{stack discipline}
for (nested) blocks.
So the use of the block statement in the second transition
axiom ensures that prior to the execution of the procedure body the
formal parameters are \emph{simultaneously} instantiated to the actual
parameters and that upon termination of a procedure call the formal
parameters are restored to their initial values.  Additionally, the
block statement limits the scope of the formal parameters so that they
are not accessible upon termination of the procedure call.  So the
second transition axiom describes the \emph{call-by-value} parameter
mechanism.

Based on the transition relation $\ra$ we consider two variants
of input/output semantics for recursive programs $S$
refering to the set $\Sigma$ of states $\sg, \tau$.
The {\it partial correctness semantics} is a mapping
$\MS{S}: \Sigma \ra {\cal P}(\Sigma)$ defined by
\[ 
 \MS{S}(\sg)=\C{\tau \mid <S,\sg> \tra <E,\tau>}. 
\]
The {\it total correctness semantics} is a mapping
$\MT{S}: \Sigma \ra {\cal P}(\Sigma \U \C{\bot} )$ defined by
\[ 
 \MT{S}(\sg) =\MS{S}(\sg) \U \C{\bot \mid S\ \mbox{can diverge from}\ \sg}. 
\]
Here $\bot$ is an error state signalling divergence, i.e.,
an infinite sequence of transitions starting in the configuration
$ <S,\sg>$.

\section{Proof systems for partial and total correctness}
\label{sec:ver}

Program correctness is expressed by
{\it correctness formulas} of the form $\HT{p}{S}{q}$,
where $S$ is a program and $p$ and $q$ are assertions. 
The assertion $p$ is the {\it precondition} of the correctness formula and $q$ is the {\it postcondition}.
A correctness formula \HT{p}{S}{q}\ is true in the sense of 
partial correctness
if every terminating computation of $S$ that starts in a state satisfying $p$ terminates in a state satisfying $q$.
And \HT{p}{S}{q}\ is true in the sense of total correctness if every
computation of $S$ that starts in a state satisfying $p$ terminates and its final state satisfies $q$.
Thus in the case of partial correctness, 
diverging computations of $S$ are not taken into account.

Using the semantics $\cal M$ and ${\cal M}_{tot}$, we formalize 
these two interpretations of correctness formulas 
uniformly as set theoretic inclusions as follows
(cf.~\cite{ABO09}).
For an assertion $p$ let $\B{p}$ denote the set of states
satisfying $p$. Then we define:

\begin{enumerate}\smallromani
\item The correctness formula \HT{p}{S}{q}\ is true in the
sense of {\it partial correctness}, 
abbreviated by $\Mo \HT{p}{S}{q}$, if $\MS{S}(\B{p}) \sse \B{q}$.
\vspace{2mm}

\item The correctness formula \HT{p}{S}{q}\ is true in the
sense of {\it total correctness}, 
abbreviated by $\Mt \HT{p}{S}{q}$, if $\MT{S}(\B{p}) \sse \B{q}$.
\end{enumerate}
Since by definition $\bot \not \in \B{q}$, part (ii) indeed formalizes the above intuition about total correctness.



\subsection*{Partial Correctness} 
\label{subsec:rec2:pc}

Partial correctness of {\bf while} programs is proven using the
customary proof system {\it PD} consisting of the group of axioms and
rules \myref{rul:skip}--\myref{rul:cons} shown in the appendix.
Consider now partial correctness of recursive programs.  First, we
introduce the following rule that deals with the block statement. 

\III

\NI
BLOCK \mylabel{rul:block} 
\[
\frac{\HT{p}{\bar{x} := \bar{t}; S}{q}}
{\HT{p}{\block{\local \bar{x} := \bar{t}; S}}{q}}
\]
where $var(\bar{x}) \cap free(q) = \ES$.
\III

By $free(q)$ we denote here the set of all free simple and array
variables that have a free occurrence in the assertion $q$.

The main issue is how to deal with the procedure calls. To this end, we
want to adjust the proofs of `generic' procedure calls to arbitrary
ones.  The definition of a generic call and the conditions for the
correctness of such an adjustment process refer to the assumed set of
procedure declarations $D$.  By a generic call of a procedure $P$ we
mean a call of the form $P(\bar{x})$, where $\bar{x}$ is a sequence of
fresh (w.r.t.~$D$) variables.

First, we extend the definition of $change(S)$ to recursive programs
and sets of procedure declarations as follows:
\[
\begin{array}{l}
change(\block{\local \bar{x}:=\bar{t}; S}) = change(S) \setminus \C{\bar{x}}, \\

change(P(\bar{u}) :: S) = change(S) \setminus \C{\bar{u}}, \\
change(\{P(\bar{u}) :: S\} \cup D) = change(P(\bar{u}) :: S) \cup change(D), \\
change(P(\bar{t})) = \ES.
\end{array}
\]

The adjustment of the generic procedure calls is then taken care of by the following proof rule
that refers to the set of procedure declarations $D$:
\III

\NI
INSTANTIATION \mylabel{rul:inst}
\[
\frac{\HT{p}{P(\bar{x})}{q}}
{\HT{p[\bar{x}:=\bar{t}]}{P(\bar{t})}{q[\bar{x}:=\bar{t}]}}
\]
where $var(\bar{x}) \cap var(D) = 
var(\bar{t}) \cap change (D) = \ES$ and $P(\bar{u}) :: S \in D$ for some $S$.
\III

In the following rule for recursive procedures with parameters
we use the provability symbol $\vdash$ to
refer to the proof system \emph{PD} augmented with the
auxiliary axiom and rules \ref{rul:inv1}--\ref{rul:sub}
defined in the appendix and the above two proof rules.  

\III

\NI
RECURSION \mylabel{rul:rec3} 
\[
\begin{array}{l}
\HT{p_1}{P_1(\bar{x}_1)}{q_1},\ldots,\HT{p_n}{P_n(\bar{x}_n)}{q_n} \vdash \HT{p}{S}{q},                    \\
\HT{p_1}{P_1(\bar{x}_1)}{q_1},\ldots,\HT{p_n}{P_n(\bar{x}_n)}{q_n} \vdash \\
\qquad \HT{p_i}{\block{\local \bar{u}_i:=\bar{x}_i; S_i}}{q_i}, \ i \in \{1, \LL, n\} \\
[-\medskipamount]
\hrulefill                                                      \\
\HT{p}{S}{q} 
\end{array}
\]
where $D = P_1(\bar{u}_1) ::S_1,\ldots,P_n(\bar{u}_n) ::S_n$ 
and $var(\bar{x}_i) \cap var(D) = \ES$ for $i \in \{1, \LL, n\}$.

\III

The intuition behind this rule is as follows. Say that a program $S$
is $(p,q)$-\emph{correct} if $\HT{p}{S}{q}$ holds in the sense of
partial correctness.  The second premise of the rule states that we
can establish from the \emph{assumption} of the
$(p_i,q_i)$-correctness of the `generic' procedure calls
$P_i(\bar{x}_i)$ for $i \in \{1, \ldots, n\}$, the
$(p_i,q_i)$-correctness of the procedure bodies $S_i$ for $i \in \{1,
\ldots, n\}$, which are adjusted as in the transition axiom that deals
with the procedure calls. Then we can prove the
$(p_i,q_i)$-correctness of the procedure calls $P_i(\bar{x}_i)$
unconditionally, and thanks to the first premise establish the
$(p,q)$-correctness of the recursive program $S$.

To prove {\it p}artial correctness of 
{\it r}ecursive programs with parameters
we use the  proof system {\it PR\/} 
that is  obtained by extending the proof system
{\it PD} by the block rule, 
the instantiation rule, 
the recursion rule,
and the auxiliary axiom and rules \ref{rul:inv1}--\ref{rul:sub}.

Note that when we deal only with one recursive procedure and use the procedure
call as the considered recursive program, this rule
simplifies to
\[ 
  \frac{\HT{p}{P(\bar{x})}{q}  \vdash \HT{p}{\block{\local \bar{u}:=\bar{x}; S}}{q}}
        { \HT{p}{P(\bar{x})}{q}                          }
\]
where $D = P(\bar{u}) ::S$ and $var(\bar{x}) \cap var(D) = \ES$.

\subsection*{Total Correctness} 
\label{subsec:rec-tot}

Total correctness of {\bf while} programs is proven using
the proof system {\it TD} consisting of the group of axioms
and rules  
\myref{rul:skip}--\myref{rul:cond}, \myref{rul:cons}, and \myref{rul:loop2}
shown in the appendix.
For total correctness of recursive programs we need
a modification of the recursion rule.
The provability symbol $\vdash$ refers now to
the proof system \emph{TD} augmented with 
the auxiliary rules \ref{rul:disj}--\ref{rul:sub},
the block rule and the instantiation rule.
The proof rule is a minor variation of a rule originally proposed in \cite{AB90} and
has the following form:
\III

\NI
RECURSION II \mylabel{rul:rec4}
\[
\begin{array}{l}
\HT{p_1}{P_1(\bar{x}_1)}{q_1},\ldots,\HT{p_n}{P_n(\bar{x}_n)}{q_n} \vdash \HT{p}{S}{q},                    \\
\HT{p_1\wedge t< z}{P_1(\bar{x}_1)}{q_1},\ldots,\HT{p_n \wedge t< z}{P_n(\bar{x}_n)}{q_n} \vdash \\
\qquad \HT{p_i\wedge t= z}{\block{\local \bar{u}_i:=\bar{x}_i; S_i}}{q_i}, \ i \in \{1, \LL, n\} \\
[-\medskipamount]
\hrulefill                                                      \\
\HT{p}{S}{q} 
\end{array}
\]
where $D = P_1(\bar{u}_1) ::S_1,\ldots,P_n(\bar{u}_n) ::S_n$, 
$var(\bar{x}_i) \cap var(D) = \ES$ for $i \in \{1, \LL, n\}$,
and $z$ is an integer variable that does not occur in
$p_i, t, q_i$ and $S_i$ for $i \in \{1, \LL, n\}$ and
is treated in the proofs as a constant, which means that in these
proofs neither the $\te$-introduction rule \ref{rul:intro} nor the substitution rule \ref{rul:sub}
defined in the appendix is applied to $z$.
\III

To prove {\it t}otal correctness of 
{\it r}ecursive programs with parameters
we use the proof system {\it TR\/} 
that is obtained by extending the proof system
{\it TD} by the block rule,
the instantiation rule,
the recursion rule II,
and the auxiliary rules \ref{rul:disj}--\ref{rul:sub}.

As before, in the case of one recursive procedure this rule can be
simplified to
\[
 \begin{array}{l}
  \HT{p\wedge t< z}{P(\bar{x})}{q} \vdash 
  \HT{p\wedge t= z}{\block{\local \bar{u}:=\bar{x};  S}}{q}, \\
  p\rightarrow t\geq  0                     \\
  [-\medskipamount]
  \hrulefill                                                      \\
  \HT{p}{P(\bar{x})}{q} 
 \end{array}
\]
where $D = P(\bar{u}) ::S$, $var(\bar{x}) \cap var(D) = \ES$
and $z$ is an integer variable that does not occur in
$p, t, q$ and $S$ and
is treated in the proof as a constant.

\section{Modularity}
\label{sec:mod}

Proof system {\it TR} allows us to establish 
total correctness of recursive programs directly.
However, sometimes it is more convenient to decompose 
the proof of total correctness into two separate proofs, one of
partial correctness and one of termination.  
More specifically, given a correctness formula $\HT{p}{S}{q}$, 
we first establish its partial correctness, 
using proof system {\it PR}. 
Then, to show termination it suffices to prove the simpler 
correctness formula $\HT{p}{S}{\T}$ using proof system {\it TR}.

These two different proofs can be combined into one using the
following general proof rule for total correctness:
\III

\NI
DECOMPOSITION \mylabel{rul:decomp}
\[
\begin{array}{l}
\vdash_{PR} \HT{p}{S}{q},      \\
\vdash_{TR} \HT{p}{S}{\T}     \\
[-\medskipamount]
\hrulefill                              \\
\HT{p}{S}{q}
\end{array}
\]
where $\vdash_{PR}$ and $\vdash_{PR}$ refer to the proofs in the
proof systems {\it PR} and {\it TR}, respectively.
\III

The decomposition rule and other auxiliary rules
like \myref{rul:disj} or \myref{rul:conj} allow us to combine
two correctness formulas derived \emph{independently}.
In some situations it is helpful to reason about procedure calls in a hierarchical way, by
first deriving one correctness formula and then using it in a proof of 
another correctness formula. The following modification of the above simplified version
of the recursion rule illustrates this principle, where we limit ourselves to a two-stage proof
and one procedure:
\III

\NI
MODULARITY
\[
\begin{array}{l}
\HT{p_0}{P(\bar{x})}{q_0}  \vdash \HT{p_0}{\block{\local \bar{u}:=\bar{x}; S}}{q_0}, \\
\HT{p_0}{P(\bar{x})}{q_0}, \HT{p}{P(\bar{x})}{q}  \vdash \HT{p}{\block{\local \bar{u}:=\bar{x}; S}}{q} \\
[-\medskipamount]
\hrulefill                                                      \\
\HT{p}{P(\bar{x})}{q} 
\end{array}
\]
where $D = P(\bar{u}) ::S$ and $var(\bar{x}) \cap var(D) = \ES$.
                         
\III

So first we derive an auxiliary property, $\HT{p_0}{P(\bar{x})}{q_0}$
that we subsequently use in the proof of the `main' property,
$\HT{p}{P(\bar{x})}{q}$.  In general, more procedures may be used and
an arbitrary `chain' of auxiliary properties may be constructed.  In
the next section we show that such a modular approach can lead to
better structured correctness proofs.

\section{Correctness proof of the $Quicksort$ procedure}
\label{sec:quick}

We now apply the modular proof method to verify
total correctness of the \emph{Quicksort} algorithm,
originally introduced in \cite{Hoa62}. 
For a given array $a$ of type ${\bf integer} \ra {\bf integer}$
and integers $x$ and $y$ this algorithm sorts the
section $a[x:y]$ consisting of all elements $a[i]$ with $x \leq i \leq y$.
Sorting is accomplished `in situ', i.e., the elements of the
initial (unsorted) array section are permuted to achieve the sorting property.
We consider here the following version of \emph{Quicksort} close
to the one studied in \cite{FH71}.
It consists of a recursive procedure $Quicksort(m,n)$, 
where the formal parameters $\m, \n$ and the local variables $v, w$ 
are all of type ${\bf integer}$:
\VV

\begin{mytabbing}
\qquad\qquad $Quicksort(\m,\n)::$ \\
\qquad\qquad\qquad \= \IF $\m < \n$ \\
\> \textbf{then} \= $Partition(\m,\n)$;\\
\> \> $\BEGIN$\\
\> \> $\local v,w:=\ri,\lf$;\\
\> \> $Quicksort(\m,v)$;\\
\> \> $Quicksort(w,\n)$\\
\> \> $\END$\\
\> \FI
\end{mytabbing}
$Quicksort$ calls a non-recursive procedure $Partition(m,n)$ which
partitions the array $a$ suitably, using global variables $\ri,\lf,\p$ of 
type ${\bf integer}$ standing for \emph{pivot}, \emph{left}, and
\emph{right} elements:
\begin{mytabbing}
\qquad\qquad $Partition(\m,\n)::$ \\
\qquad\qquad\qquad \= $\p := a[\m];$ \\
\> $\lf,\ri := \m,\n;$\\
\> \WHILE $\lf \leq \ri$ \DO \\
\>  \quad \= \WHILE $a[\lf] < \p$ \DO $\lf:=\lf+1$ \OD; \\
\>        \> \WHILE $\p < a[\ri]$ \DO $\ri:=\ri-1$ \OD; \\
\>        \> \IF $\lf \leq \ri$ \THEN \\
\>        \> \ \ \= $swap(a[\lf],a[\ri]);$ \\
\>        \>     \> $\lf,\ri := \lf+1, \ri-1$ \\
\>        \> \FI \\
\> \OD 
\end{mytabbing}
Here for two given simple or subscripted variables $u$ and $v$ the
program $swap(u,v)$ is used to {\it swap} the values of $u$ and $v$.
So we stipulate that the following correctness formula
\[
 \HT{x = u \A y = v}{swap(u,v)}{x = v \A y = u}
\]
holds in the sense of partial and total correctness,
where $x$ and $y$ are fresh variables.

In the following $D$ denotes the set of the above two
procedure declarations and $S_Q$ the body of the
procedure $Quicksort(m,n)$.

\subsection*{Formal Problem Specification}

Correctness of $Quicksort$ amounts to proving that upon termination of the 
procedure call $Quicksort(\m,\n)$ the array section $a[\m:\n]$ is sorted and is a permutation of the input section. 
To write the desired correctness formula we introduce some notation.
The assertion
\[
 sorted(a[x:y]) \equiv 
 \forall i, j : (x \leq i\leq j \leq y \rightarrow 
 a[i]\leq a[j])
\]
states that the integer array section $a[x:y]$ is sorted.
To express the permutation property we use an auxiliary array $a_0$ 
in the section $a_0[x:y]$ of which we record the initial values of $a[x:y]$.
The abbreviation
\[
 bij(f,x,y)\ \equiv\  f \mbox{ is a bijection on }\, \integer\ 
                     \land\ 
                     \fa\, i \not\in [x:y]: f(i)=i
\]
states that $f$ is a bijection on $\integer$ which is the identity
outside the interval $[x:y]$. 
Hence 
\[
perm(a, a_0,[x:y])\ \equiv\ \exists f : (bij(f,x,y) \A \fa i  : a[i] = a_0[f(i)])
\]
specifies that the array section $a[x:y]$ is a permutation of the array section $a_0[x:y]$ and that $a$ and $a_0$ are the same elsewhere.

We can now express the correctness of $Quicksort$ by means of the following 
correctness formula:
\begin{description}
\item[\textbf{Q1}] 
$\HT{a = a_0}{Quicksort(x,y)}{perm(a, a_0,[x:y]) \A sorted(a[x:y])}$.
\end{description}
To prove correctness of $Quicksort$ in the sense of partial
correctness we proceed in stages and follow the methodology explained
in Section \ref{sec:mod}.  In other words, we establish some
auxiliary correctness formulas first, using among others the recursion
rule. Then we use them as premises in order to
derive other correctness formulas, also using the recursion rule.

\subsection*{Properties of $Partition$}
\label{subsec:partition}

In the proofs we shall use a number of properties of the $Partition$
procedure.  This procedure is non-recursive, so to verify them it
suffices to prove the corresponding properties of the procedure body
using the proof systems {\it PD} and {\it TD}, a task we leave to Nissim Francez.

More precisly, we assume the following properties of $Partition$ 
in the sense of partial correctness:
\begin{align*}
\textbf{P1}\quad  & \HT{\T}
                    {\ Partition(\m,\n)} 
                    {\ri \leq \n \A \m \leq \lf},\\[2mm]
\textbf{P2}\quad  & \C{x' \leq \m \A \n \leq y'\ \A\ perm(a,a_0,[x':y'])}\\
                  & Partition(\m,\n) \\
                  & \C{x' \leq \m \A \n \leq y'\ \A\ perm(a,a_0,[x':y'])}, \\[2mm]
\textbf{P3}\quad  & \C{\T} \\
                  & Partition(\m,\n) \\
                  & \{\ \lf > \ri \A \\
                  & \ \ (\,\fa\, i \in [\m:\ri]: \ a[i] \leq \p) \A \\
                  & \ \ (\,\fa\, i \in [\ri+1:\lf-1]: \ a[i] = \p) \A \\
                  & \ \ (\,\fa\, i \in [\lf:\n]:  \ \p \leq a[i])\},
\intertext{and the following property in the sense of total correctness:}
\textbf{P4}\quad  & \C{\m < \n}\\
                  & Partition(\m,\n) \\
                  & \C{\ri-\m < \n-\m \A \n-\lf < \n-\m }.
\end{align*}
Property \textbf{P1} states the bounds for $\ri$ and $ \lf$. 
We remark that $\lf \leq \n$ and $\m \leq \ri$ need not hold
upon termination.
Property \textbf{P2} implies
that the call $Partition(n,k)$ permutes the array section $a[\m:\n]$ and
leaves other elements of $a$ intact, but actually is a stronger
statement involving an interval $[x':y']$ that includes $[\m:\n]$, so
that we can carry out the reasoning about the recursive calls of
$Quicksort$.  Finally, property \textbf{P3} captures the main effect
of the call $Partition(n,k)$: the elements of the section $a[\m:\n]$ are
rearranged into three parts, those smaller than $\p$ (namely $a[\m:\ri]$), 
those equal to $\p$ (namely $a[\ri+1:\lf-1]$), and
those larger than $\p$ (namely $a[\lf:\n]$).
Property~\textbf{P4} is needed in the termination proof of
the $Quicksort$ procedure: it states that the subsections
$a[\m:\ri]$ and $a[\lf:\n]$ are strictly smaller than the section $a[\m:\n]$.

\subsection*{Auxiliary proof: permutation property}

In the remainder of this section we use the following abbreviation:
\[
J \equiv \m = x \A \n = y.
\]
We  first extend the permutation property \textbf{P2} to the 
procedure $Quicksort$:
\begin{align*}
\textbf{Q2} \quad & \C{perm(a, a_0,[x':y'])\wedge x'\leq x\A  y\leq y'} \\
                  & Quicksort(x,y)\\
                  & \C{perm(a, a_0,[x':y'])}
\end{align*}
Until further notice the provability symbol $\vdash$ refers
to the proof system \emph{PD} augmented with the
the block rule, the instantiation rule and the
auxiliary rules \ref{rul:disj}--\ref{rul:sub}.

The appropriate claim needed for the application of the recursion rule is:
\III

\NI
\textbf{Claim 1.}
\[
\begin{array}{ll} 
\textbf{P1}, \textbf{P2},\textbf{Q2} \vdash 
  & \C{perm(a, a_0,[x':y'])\wedge x'\leq x< y\leq y'}\\
  & \block{\local \m,\n:= x,y; S_Q}\\
  & \C{perm(a, a_0,[x':y'])}.
\end{array}
\]

\Proof\ 
In Figure~\ref{fig:aux2} a proof outline is given that uses as
assumptions the correctness formulas \textbf{P1}, \textbf{P2}, 
and \textbf{Q2}.
More specifically, the used correctness formula about the call of
$Partition$ is derived from \textbf{P1} and \textbf{P2} by the
conjunction rule. In turn, the correctness formulas about the
recursive calls of $Quicksort$ are derived from \textbf{Q2} by an
application of the instantiation rule and the invariance rule.  This
concludes the proof of Claim 1.  
\HB

\III

\begin{figure}[htb]
\normalsize
\begin{mytabbing}
\qquad\qquad 
\= \C{perm(a, a_0,[x':y'])\A x'\leq x \A  y\leq y'}                                 \\
\> $\BEGIN \ \local$ \\
\> \C{perm(a, a_0,[x':y'])\A x'\leq x \A  y\leq y'}                                 \\
\> $\m,\n:= x,y$; \\
\> \C{perm(a, a_0,[x':y'])\A x'\leq x \A y\leq y' \A J}                                 \\
\> \C{perm(a, a_0,[x':y'])\A x'\leq \m\ \A   \n\leq y'}                                  \\
\> \textbf{if} $\m < \n$ \textbf{then}                         \\
\> \qquad \C{perm(a, a_0,[x':y'])\A x'\leq \m\ \A   \n\leq y'}                                 \\
\> \qquad $Partition(\m,\n)$;                               \\
\> \qquad \C{perm(a, a_0,[x':y'])\A x'\leq \m \A \n\leq y' \A \ri\leq \n \A \m\leq \lf}                           \\
\> \qquad $\BEGIN \ \local$\\
\> \qquad \C{perm(a, a_0,[x':y'])\A x'\leq \m \A  \n\leq y' \A \ri\leq \n \A \m\leq \lf}                          \\
\> \qquad $v,w:=\ri,\lf$; \\
\> \qquad \C{perm(a, a_0,[x':y'])\A x'\leq \m \A \n\leq y' \A v\leq \n \A \m\leq w}                             \\
\> \qquad \C{perm(a, a_0,[x':y'])\A x'\leq \m \A v \leq y'  \A x'\leq w \A \n \leq y'}                           \\
\> \qquad $Quicksort(\m,v)$;\\
\> \qquad \C{perm(a, a_0,[x':y'])\A x'\leq w \A \n \leq y'} \\
\> \qquad $Quicksort(w,\n)$\\
\> \qquad \C{perm(a, a_0,[x':y']) } \\
\> \qquad \textbf{\END}                                         \\
\> \qquad \C{perm(a, a_0,[x':y'])}  \\
\> \FI                        \\
\> \C{perm(a, a_0,[x':y'])}  \\
\> \textbf{\END}                                         \\
\> \C{perm(a, a_0,[x':y'])}
\end{mytabbing}
\caption{\label{fig:aux2}Proof outline showing permutation property \textbf{Q2}.}
\end{figure}

We can now derive \textbf{Q2} by the recursion rule.
In summary, we proved 
\[
\textbf{P1}, \ \textbf{P2} \vdash \textbf{Q2}.
\]

\subsection*{Auxiliary proof: sorting property}

We can now verify that the call $Quicksort(x,y)$ sorts the array section $a[x:y]$, so
\begin{description}
\item[\textbf{Q3}]
$\HT{\T}{Quicksort(x,y)}{sorted(a[x:y])}$.
\end{description}
The appropriate claim needed for the application of the recursion rule is:

\III

\NI
\textbf{Claim 2.}
\[
\begin{array}{l} 
\textbf{P3},\ \textbf{Q2}, \ \textbf{Q3} \vdash 
\HT{\T}{\block{\local \m,\n:= x,y; S_Q}}{sorted(a[x:y])}.
\end{array}
\]

\Proof\
In Figure~\ref{fig:qs-p2} a proof outline is given that uses as
assumptions the correctness formulas \textbf{P3}, \textbf{Q2}, 
and \textbf{Q3}. 
\begin{figure}[htb]
\normalsize
\begin{mytabbing}
\qquad\qquad
\= \C{\T}                                 \\
\> $\BEGIN \ \local$ \\
\> \C{\T}                                 \\
\> $\m,\n:= x,y$; \\
\> \C{J}                                 \\
\> \textbf{if} $\m < \n$ \textbf{then}                 \\
\> \qquad \C{J \A \m < \n}                                 \\
\> \qquad $Partition(\m,\n)$;                               \\
\> \qquad \C{J \A K[v,w:=\ri,\lf]}                           \\
\> \qquad $\BEGIN \ \local$\\
\> \qquad \C{J \A K[v,w:=\ri,\lf]}                           \\
\> \qquad $v,w:=\ri,\lf$; \\
\> \qquad \C{J \A K}                           \\
\> \qquad $Quicksort(\m,v)$;\\
\> \qquad \C{sorted(a[\m:v]) \A J \A K} \\
\> \qquad $Quicksort(w,\n)$\\
\> \qquad \C{sorted(a[\m:v] \A sorted(a[w:\n] \A J \A K} \\
\> \qquad \C{sorted(a[x:v] \A sorted(a[w:y] \A K[\m,\n:=x,y]} \\
\> \qquad \C{sorted(a[x:y])}  \\
\> \qquad \textbf{\END}                                         \\
\> \qquad \C{sorted(a[x:y])}  \\
\> \FI                        \\
\> \C{sorted(a[x:y])}  \\
\> \textbf{\END}                                         \\
\> \C{sorted(a[x:y])}
\end{mytabbing}
\caption{\label{fig:qs-p2}Proof outline showing sorting property \textbf{Q3}.}
\end{figure}
In the following we justify the correctness
formulas about $Partition$ and the recursive calls of $Quicksort$
used in this proof outline.
In the postcondition of $Partition$ we use the following abbreviation:
\[
\begin{array}{lll}
  K \equiv & \ v < w \A \\
           & (\,\fa\, i \in [\m:v]: \ a[i] \leq \p) \A \\
           & (\,\fa\, i \in [v+1:w-1]: \ a[i] = \p) \A \\
           & (\,\fa\, i \in [w:\n]:  \ \p \leq a[i]).
\end{array}
\]
Observe that the correctness formula 
\[
\HT{J}{Partition(\m,\n)}{J \A K[v,w:=\ri,\lf]}
\]
is derived from \textbf{P3} by the invariance rule.
Next we verify the correctness formulas
\begin{equation}
  \label{equ:inner1}
\C{J \A K}                           \\
Quicksort(\m,v) \\
\C{sorted(a[\m:v])\A J \A K},
\end{equation}
and
\begin{equation}
  \label{equ:inner2}
\begin{array}{l}
\C{sorted(a[\m:v])\A J \A K} \\
Quicksort(w,\n) \\
\C{sorted(a[\m:v] \A sorted(a[w:\n] \A J \A K}.
\end{array}
\end{equation}
about the recursive calls of $Quicksort$.

\III

\NI
\emph{Proof of (\ref{equ:inner1})}. 
By applying the instantiation rule to \textbf{Q3}, we obtain
\begin{description}
\item[\textbf{A1}]  $\HT{\T}{Quicksort(\m,v)}{sorted(a[\m:v])}$.
\end{description}
Moreover, by the invariance axiom, we have
\begin{description}
\item[\textbf{A2}]  $\HT{J}{Quicksort(\m,v)}{J}$.
\end{description}
By applying the instantiation rule to \textbf{Q2}, we then obtain
\begin{align*}
  &\C{perm(a,a_0,[x':y'])\A x'\leq \m \A v \leq y'} \\
  & Quicksort(\m,v) \\
  &\C{perm(a,a_0,[x':y'])}.
\end{align*}
Applying next the substitution rule with the substitution $[x',y' := \m,v]$
yields
\begin{align*}
  &\C{perm(a,a_0,[\m:v])\A \m \leq \m \A v\leq v} \\
  & Quicksort(\m,v) \\
  &\C{perm(a,a_0,[\m:v])}.
\end{align*}
So by a trivial application of the consequence rule, we obtain
\[
\HT{a=a_0}{Quicksort(\m,v)}{perm(a,a_0,[\m:v])}.
\]
We then obtain by an application of the invariance rule 
\[
\HT{a=a_0 \A K[a:=a_0]}{Quicksort(\m,v)}{perm(a, a_0,[\m:v]) \A K[a:=a_0]}.
\]
Note now the following implications:
\[
\begin{array}{l}
 K \ra \te a_0: (a= a_0 \A K[a:= a_0]), \\[1mm]
 perm(a, a_0,[\m:v]) \A K[a:= a_0] \ra K.
\end{array}
\]
So we conclude 
\begin{description}
\item[\textbf{A3}]  $\HT{K}{Quicksort(\m,v)}{K}$
\end{description}
by the $\te$-introduction rule and the consequence rule.
Combining the correctness formulas \textbf{A1}--\textbf{A3} 
by the conjunction rule we get (\ref{equ:inner1}).
\III

\NI
\emph{Proof of (\ref{equ:inner2})}. 
In a similar way as above, we can prove the correctness formula
\[
\HT{a=a_0}{Quicksort(w,\n)}{perm(a,a_0,[w: \n])}.
\]
By an application of the invariance rule we obtain
\[
\begin{array}{l}
\C{a=a_0 \A sorted(a_0[\m:v]) \A v < w} \\
Quicksort(w,\n) \\
\C{perm(a, a_0,[w: \n]) \A sorted(a_0[\m:v]) \A v < w}.
\end{array}
\]
Note now the following implications:
\[
\begin{array}{l}
v < w \A sorted(a[\m:v]) \ra 
\te a_0: (a= a_0 \A sorted(a_0[\m:v]) \A v < w), \\[1mm]
(perm(a, a_0,[w:\n]) \A sorted(a_0[\m:v]) \A v < w) \ra sorted(a[\m:v]).
\end{array}
\]
So we conclude 
\begin{description}
\item[\textbf{B1}]
$\HT{v<w \A sorted(a[\m:v])}{Quicksort(w,\n)}{sorted(a[\m:v])}$
\end{description}
by the $\te$-introduction rule and the consequence rule.
Further, by applying the instantiation rule to \textbf{Q3} we obtain
\begin{description}
\item[\textbf{B2}]
$\HT{\T}{Quicksort(w,\n)}{sorted(a[w:\n])}$.
\end{description}
Next, by the invariance axiom we obtain
\begin{description}
\item[\textbf{B3}]
$\HT{J}{Quicksort(w,\m)}{J}$.  
\end{description}
Further, using the implications
\[
 \begin{array}{l}
   K \ra \te a_0: (a= a_0 \A K[a:= a_0]), \\[1mm]
   perm(a, a_0,[w:\n]) \A K[a:= a_0] \ra K,
 \end{array}
\]
we can derive from \textbf{Q2}, in a similar manner as in the proof of \textbf{A3},
\begin{description}
\item[\textbf{B4}]
$\HT{K}{Quicksort(w,\n)}{K}$.  
\end{description}
Combining the correctness formulas \textbf{B1}--\textbf{B4} by the conjunction rule and observing that $K \ra v<w$ holds, we get
(\ref{equ:inner2}).

\III

The final application of the consequence rule in the proof outline given in Figure \ref{fig:qs-p2}
is justified by the following crucial implication:
\begin{align*}
& sorted(a[x:v]) \A sorted(a[w:y]) \A K[\m,\n:=x,y] \ra \\
& sorted(a[x:y]).
\end{align*}
Also note that $J \A \m \geq \n \ra sorted(a[x:y])$, so the implicit \textbf{else} branch is properly taken care of.
This concludes the proof of Claim 2.
\HB

\III

We can now derive \textbf{Q3} by the recursion rule.
In summary, we proved 
\[
\textbf{P3},  \ \textbf{Q2} \vdash \textbf{Q3}.
\]
The proof of partial correctness of $Quicksort$ is now immediate: it
suffices to combine \textbf{Q2} and \textbf{Q3} by the conjunction
rule.  Then after applying the substitution rule with the substitution
$[x',y' := x,y]$ and the consequence rule we obtain \textbf{Q1},
or more precisely
\[
\textbf{P1}, \ \textbf{P2}, \ \textbf{P3} \vdash \textbf{Q1}.
\]

\subsection*{Total Correctness}

To prove termination, 
by the decomposition rule discussed in Section \ref{sec:mod},
it suffices to establish 
\begin{description}
\item[\textbf{Q4}]  $\HT{\T}{Quicksort(x,y)}{\T}$
\end{description}
in the sense of total correctness. In the proof we rely on
the property \textbf{P4} of $Partition$:
\[
 \HT{\m<\n}{Partition(\m,\n)}{\ri-\m < \n-\m \A \n-\lf < \n-\m}.
\]

The provability symbol $\vdash$ refers below to
the proof system \emph{TD} augmented with 
the block rule, the instantiation rule and the
the auxiliary rules \ref{rul:disj}--\ref{rul:sub}.
For the termination proof of the recursive procedure call $Quicksort(x,y)$
we use 
\[  
  t \equiv \max(y-x,0)
\]
as the bound function.
Since $t \geq 0$ holds, the appropriate claim needed for the application of the 
recursion rule II is:

\III

\NI
\textbf{Claim 3.}
\[
\begin{array}{l}
  \textbf{P4},\ \HT{t<z}{Quicksort(x,y)}{\T}  \vdash \\
  \HT{t=z}{\block{\local \m,\n:= x,y; S_Q}}{\T}.
\end{array}
\]

\Proof\
In Figure~\ref{fig:qs-t} a proof outline for total correctness
is given that uses as assumptions the correctness formulas 
\textbf{P4} and $\HT{t<z}{Quicksort(x,y)}{\T}$.
\begin{figure}[htb]
\normalsize
\begin{mytabbing}
\qquad\qquad 
\= \C{t=z}                                 \\
\> $\BEGIN \ \local$ \\
\> \C{\max(y-x,0)=z}                                 \\
\> $\m,\n:= x,y$; \\
\> \C{\max(\n-\m,0)=z}                          \\
\> \textbf{if} $n<k$ \textbf{then}              \\
\> \qquad \C{\max(\n-\m,0)=z \A \m < \n}        \\
\> \qquad \C{\n-\m=z \A \m < \n}                \\
\> \qquad $Partition(\m,\n)$;                               \\
\> \qquad \C{\n-\m=z \A \m < \n \A \ri-\m < \n-\m \A \n-\lf < \n-\m}   \\
\> \qquad $\BEGIN \ \local$\\
\> \qquad \C{\n-\m=z \A \m < \n \A \ri-\m < \n-\m \A \n-\lf < \n-\m}   \\
\> \qquad $v,w:=\ri,\lf$;\\
\> \qquad \C{\n-\m=z \A \m < \n \A v-\m < \n-\m \A \n-w < \n-\m}       \\
\> \qquad \C{\max(v-\m,0) < z \A \max(\n-w,0) < z}                     \\
\> \qquad $Quicksort(\m,v)$;\\
\> \qquad \C{\max(\n-w,0) < z}                           \\
\> \qquad $Quicksort(w,\n)$                         \\
\> \qquad \C{\T}  \\
\> \qquad \textbf{\END}                                         \\
\> \qquad \C{\T}  \\
\> \FI                        \\
\> \C{\T}  \\
\> \textbf{\END}                                         \\
\> \C{\T} 
\end{mytabbing}
\caption{\label{fig:qs-t}Proof outline establishing termination of the $Quicksort$ procedure.}
\end{figure}
In the following we justify the correctness
formulas about $Partition$ and the recursive calls of $Quicksort$
used in this proof outline.
Since $\m,\n,z \not\in change(D)$, 
we deduce from \textbf{P4}
using the invariance rule the correctness formula
\begin{equation}
\label{equ:part-t}
\begin{array}{l}
\C{\n-\m=z \A \m<\n} \\
Partition(\m,\n) \\
\C{\n-\m=z \A \ri-\m < \n-\m \A \n-\lf < \n-\m}.
\end{array}
\end{equation}
Consider now the assumption
\[
\HT{t<z}{Quicksort(x,y)}{\T}.
\]
Since $\n,w,z \not\in change(D)$, the instantiation rule 
and the invariance rule yield
\[
\begin{array}{l}
\C{\max(v-\m,0) < z \A \max(\n-w,0) < z} \\
Quicksort(\m,v) \\
\C{\max(\n-w,0) < z}
\end{array}
\]
and
\[
\HT{\max(\n-w,0) < z}{Quicksort(w,\n)}{\T}.
\]
The application of the consequence rule preceding the first recursive
call of $Quicksort$ is justified by the following two implications:
\[
\begin{array}{l}
(\n-\m=z \A \m < \n \A v-\m < \n-\m)  \ra \max(v-\m,0) < z, \\[1mm]
(\n-\m=z \A \m < \n \A \n-w < \n-\m)  \ra \max(\n-w,0) < z.
\end{array}
\]
This completes the proof of Claim 3.
\HB

\III

Applying now the simplified version of the recursion rule II we derive \textbf{Q4}.
In summary, we proved 
\[
\textbf{P4} \vdash \textbf{Q4}.
\]

\section{Conclusions}
\label{sec:conc}

The issue of modularity has been by now well-understood in the area of
program construction. It also has been addressed in the program
verification. Let us just mention two references, an early one and a
recent one: \cite{HO82} focused on modular verification of temporal
properties of concurrent programs which were modelled as a set of
modules that interact by means of procedure calls.  In turn,
\cite{Tag08} considered modular verification of heap manipulating
programs, where the focus has been on the automatic extraction and
verification specifications.

However, to our knowledge no approach has been proposed to deal with
correctness of recursive programs in a modular fashion.  When proving
correctness of the $Quicksort$ program we found that the simple
approach here proposed allowed us to structure the proof better
by establishing the `permutation property' first and then using
it in the proof of the `sorting property'.

So in our approach we propose modularity at the level of \emph{proofs}
and not at the level of \emph{programs}.  This should be of help when
organizing a mechanically verified correctness proof, by expressing
the proofs of the subsidiary properties as subsidiary lemmas.
In general, modular correctness proofs of programs are proofs from assumptions about
subprograms, which can be considered as `black boxes' of the given programs.
Zwiers \cite{Zwi89} has investigated an appropriate notion of completeness 
for such proofs from assumptions about black boxes, called
\emph{modular completeness}.

The first proof of partial correctness of \emph{Quicksort}  
is given in \cite{FH71}.
That proof establishes the permutation and the sorting property simultaneously,
in contrast to our approach.
For dealing with recursive procedures, \cite{FH71} use proof rules
corresponding to our rules for blocks, instantiation,
and recursion (for the case of one recursive procedure).
They also use a so-called \emph{adaptation rule} of \cite{Hoa71-proc}
that allows one to adapt a given correctness formula about a program 
to other pre- and postconditions. In our approach we use
several auxiliary rules which together have the same effect 
as the adaptation rule. The expressive power of the adaptation rule
has been analyzed in \cite{Old83}.
No proof rule for the termination of recursive procedures 
is proposed in \cite{FH71},
only an informal argument is given why \emph{Quicksort} terminates.
An informal proof of total correctness of \emph{Partition} is given
in \cite{Hoa71-Find} as part of the program \emph{Find} given in \cite{Hoa61}.

The recursion rule is modelled after the so-called Scott induction
rule for fixed points that appeared first in the unpublished
manuscript Scott and De Bakker \cite{SB69}.  Recursion rule II for
total correctness is taken from America and De Boer \cite{AB90}, where
also the completeness of a proof system similar to {\it TR\/} is
established.  The modularity rule corresponds to a theorem due to
Beki{\'c} \cite{Bec69} which states that for systems of monotonic functions
iterative fixed points coincide with simultaneous fixed points.

\section*{Acknowledgment}

We thank the reviewer for helpful suggestions.

\section*{Appendix}

We list here the used axioms and proof rules that were not defined earlier in the text.
To establish correctness of \textbf{while} programs we rely on the following axioms and proof rules.
In the proofs of partial correctness the loop rule is used, 
while in the proofs of total correctness the loop II rule is used.


\begin{Axiom} SKIP \mylabel{rul:skip}
\[ \HT{p}{skip}{p} \]
\end{Axiom}

\begin{Axiom} ASSIGNMENT \mylabel{rul:assi}
\[ \HT{p[u:=t]}{u:=t}{p} \]
\end{Axiom}

\begin{Axiom} PARALLEL ASSIGNMENT \mylabel{rul:assip}
\[ 
 \HT{p[\bar{x}:=\bar{t}]}{\bar{x}:=\bar{t}}{p} 
\]
\end{Axiom}

\begin{Rule} COMPOSITION \mylabel{rul:comp}
\[ \frac{ \HT{p}{S_1}{r}, \HT{r}{S_2}{q}                }
        { \HT{p}{S_1;\ S_2}{q}                          }\]
\end{Rule}

\begin{Rule} CONDITIONAL \mylabel{rul:cond}
\[ \frac{ \HT{p \A B}{S_1}{q}, \HT{p \A \neg B}{S_2}{q}         }
        { \HT{p}{\ITE{B}{S_1}{S_2}}{q}                          }\]
\end{Rule}

\begin{Rule} LOOP \mylabel{rul:loop}
\[ \frac{ \HT{p \A B}{S}{p}                             }
        { \HT{p}{\WDD{B}{S}}{p \A \neg B}               }\]
\end{Rule}

\begin{Rule} CONSEQUENCE \mylabel{rul:cons}
\[ \frac{ p \ra p_1, \HT{p_1}{S}{q_1}, q_1 \ra q        }
        { \HT{p}{S}{q}                                  }\]
\end{Rule}

\begin{Rule} LOOP II \mylabel{rul:loop2}
\[
\begin{array}{l}
\HT{p \A B}{S}{p},                      \\
\HT{p \A B \A t=z}{S}{t<z},             \\
p\ \ra t \geq 0                         \\
[-\medskipamount]
\hrulefill                              \\
\HT{p}{\WDD{B}{S}}{p \A \neg\ B}
\end{array}
\]
where $t$ is an integer expression and $z$ is an integer variable that
does not appear in $p,B,t$ or $S$.
\end{Rule}


Additionally, we rely on the following auxiliary axioms and proof rules
that occasionally refer to the assumed set of procedure declarations $D$.

\begin{AuxAxiom} INVARIANCE \mylabel{rul:inv1}
\[ \HT{p}{S}{p} 
\]
where $free(p) \I (change(D) \cup change(S)) = \ES$.
\end{AuxAxiom}

\begin{AuxRule} DISJUNCTION \mylabel{rul:disj}
\[ \frac{ \HT{p}{S}{q}, \HT{r}{S}{q}    }
        { \HT{p \Or r}{S}{q}            }\]
\end{AuxRule}

\begin{AuxRule} CONJUNCTION \mylabel{rul:conj}
\[ \frac{ \HT{p_1}{S}{q_1}, \HT{p_2}{S}{q_2}     }
        { \HT{p_1 \A p_2}{S}{q_1 \A q_2}         } \]
\end{AuxRule}

\begin{AuxRule} $\te$-INTRODUCTION \mylabel{rul:intro}
\[ \frac{ \HT{p}{S}{q}          }
        { \HT{\te x:p}{S}{q}    }
\]
where $x \not\in change(D) \cup change(S) \cup free(q)$.
\end{AuxRule}

\begin{AuxRule} INVARIANCE \mylabel{rul:inv2}
\[ \frac{ \HT{r}{S}{q}           }
        { \HT{p \A r}{S}{p \A q} } 
\]
where $free(p) \I (change(D) \cup change(S)) =\ES$.
\end{AuxRule}

\begin{AuxRule} SUBSTITUTION \mylabel{rul:sub}
\[ \frac{ \HT{p}{S}{q} }  
        { \HT{p[\bar{z}:=\bar{t}]}{S}{q[\bar{z}:=\bar{t}]} } 
\]
where $(var(\bar{z}) \cup var(\bar{t})) \I
          (change(D) \cup change(S)) =\ES$.
\end{AuxRule}

\bibliographystyle{plain}

\bibliography{apt,abo}
\end{document}